\documentclass[cup6b]{cupbook}
\usepackage{graphicx}
\usepackage [numbers]{natbib}
\title[Rome, Italy, 27--30 April 2009]
      {The coming of age of X-ray polarimetry}
\author{}
\date{}
\begin{document}
\pagenumbering{arabic}
%%%%%%%%%%%%%%%%%%%%%%%%%%%%%%%%%%%%%%%%%%%%%%%%%%%%

% YOUR CONTRIBUTION GOES HERE

% Insert here the author list and the affiliation
\author[Rishin et al.]{Rishin P.V., Biswajit Paul, Duraichelvan R.\\Raman Research Institute, Sadashivanagar, Bangalore--560080, India\and Marykutty James, Jincy Devasia\\School of Pure and Applied Physics, Mahatma Gandhi University,\\Kottayam--686560, Kerala, India\and Ramanath Cowsik\\McDonnell Center for the Space Sciences, Department of Physics,\\ Washington University, St. Louis, MO 63130, USA}

% Title of the contribution
\chapter{Development of a Thomson X-ray Polarimeter}

% Abstract
\abstract{We describe the current status of the design and development of a Thomson X-ray polarimeter suitable for a small satellite mission. Currently we are considering two detector geometries, one using rectangular detectors placed on four sides of a scattering element and the other using a single cylindrical detector with the scattering element at the center. The rectangular detector configuration has been fabricated and tested. The cylindrical detector is currently under fabrication. In order to compensate any pointing offset of the satellite, a collimator with a flat topped response has been developed that provides a constant effective area over an angular range. We have also developed a double crystal monochromator/polariser for the purpose of test and calibration of the polarimeter. Preliminary test results from the developmental activities are presented here.}

\section{Introduction}
A Thomsom X-ray polarimeter experiment has been proposed for a small satellite mission of the Indian Space Research Organization (ISRO). Currently, a laboratory model has been developed. This experiment will be suitable for X-ray polarisation measurement of hard X-ray sources like accretion powered pulsars, black hole candidates in low-hard states etc. (Kallman 2004, Weisskopf et al. 2006). For about 50 brightest X-ray sources a Minimum Detectable Polarisation of 2--3 \% will be achieved with the final configuration.

Two configurations are considered based on the geometry of the detector element, (1) Rectangular and (2) Cylindrical Detectors. The X-ray polarisation will be measured by spinning the platform around the viewing axis. In both the cases, energy range covered will be 5--30 keV. Here we describe the design and current state of development of the Thomson X-ray polarimeter and also the development of a test and calibration facility.

\section{Thomson X-ray Polarimeter Design}

% Figure: Polarimeter configuration with rectangular detectors
\begin{figure}
\centering
\includegraphics[scale=0.35]{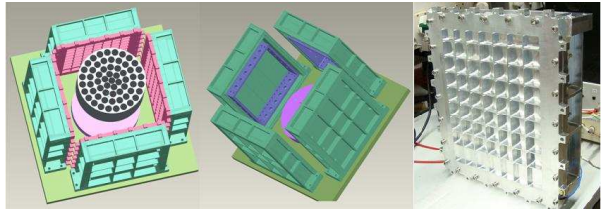}
\caption{Polarimeter configuration with rectangular detectors.}
\label{f1}
\end{figure}

{\it Rectangular Detectors:} In Figure \ref{f1} we have shown the configuration of the Thomson Polarimeter with rectangular detectors. Here the detector elements are flat multi--wire proportional counters placed on four sides of a 900 cm$^2$ disk--like scattering element made of Be or Li. Each detector also has a photon collection area of 900 cm$^2$.

%Figure: Cylindrical detector fabrication design
\begin{figure}
\centering
\includegraphics[scale=0.35]{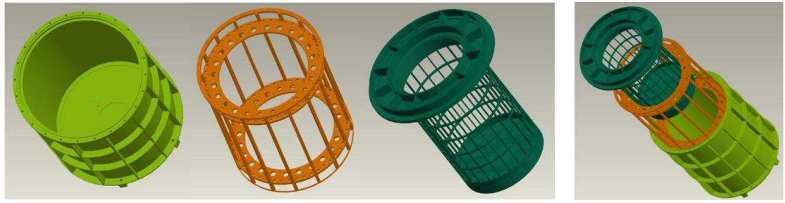}
\caption{Cylindrical detector fabrication design and the assembly.}
\label{f2}
\end{figure}

{\it Cylindrical Detectors:}
The second configuration uses cylindrical detectors as shown in Figure \ref{f2}. In this configuration, we have a 360$^\circ$ coverage of the scattered photons which is an advantage compared to the rectangular detectors. The experiment will have two such detectors, each with a photon collection area of about 350 cm$^2$. In a cylindrical detector, we expect to have uniform gain and quantum efficiency in all directions which will help to reduce systematic errors while searching for faint signal. 
 
\begin{figure}[h]
\centering
\hfill \hfill \hfill
\includegraphics[scale=0.23]{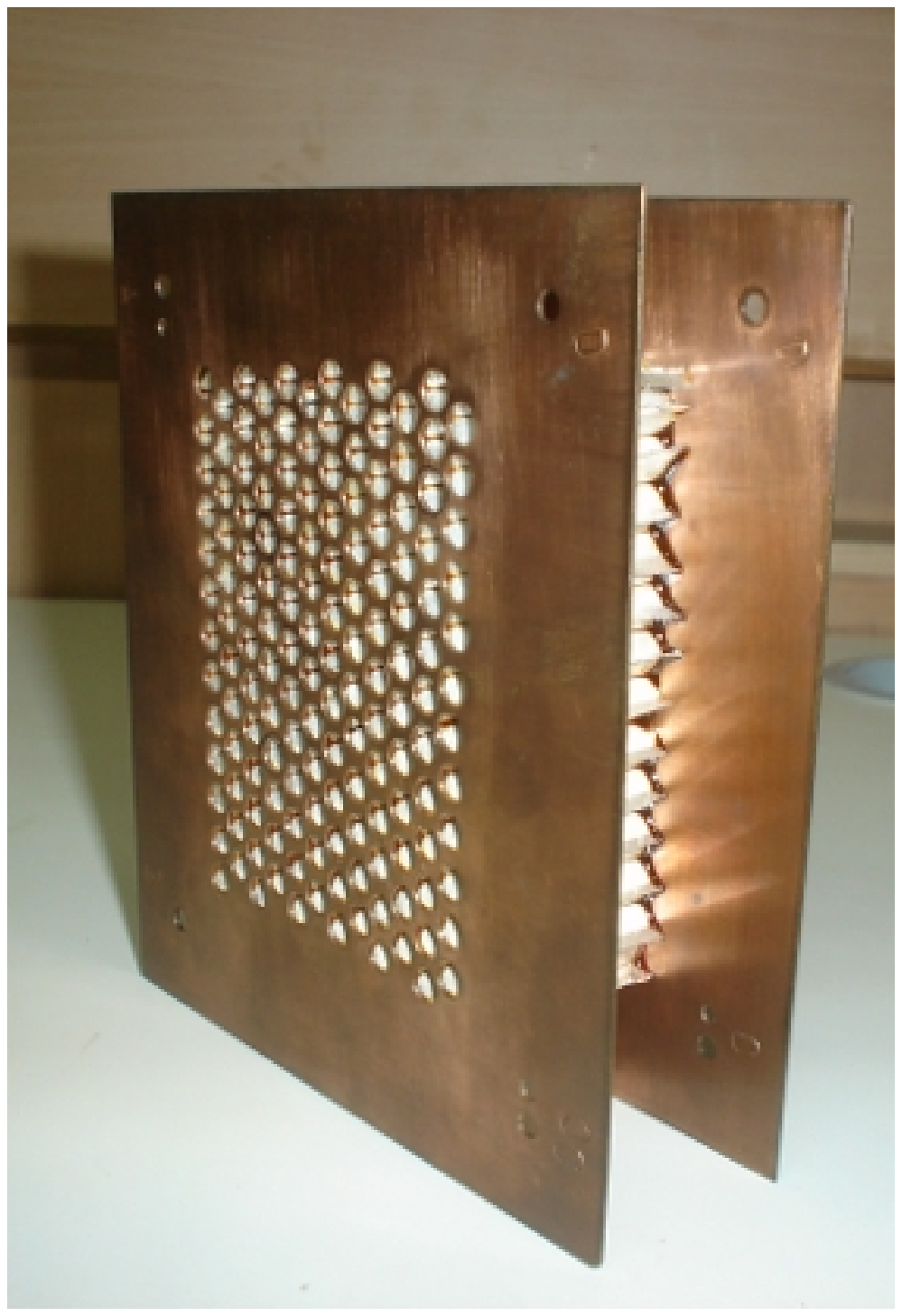}
\includegraphics[scale=0.45]{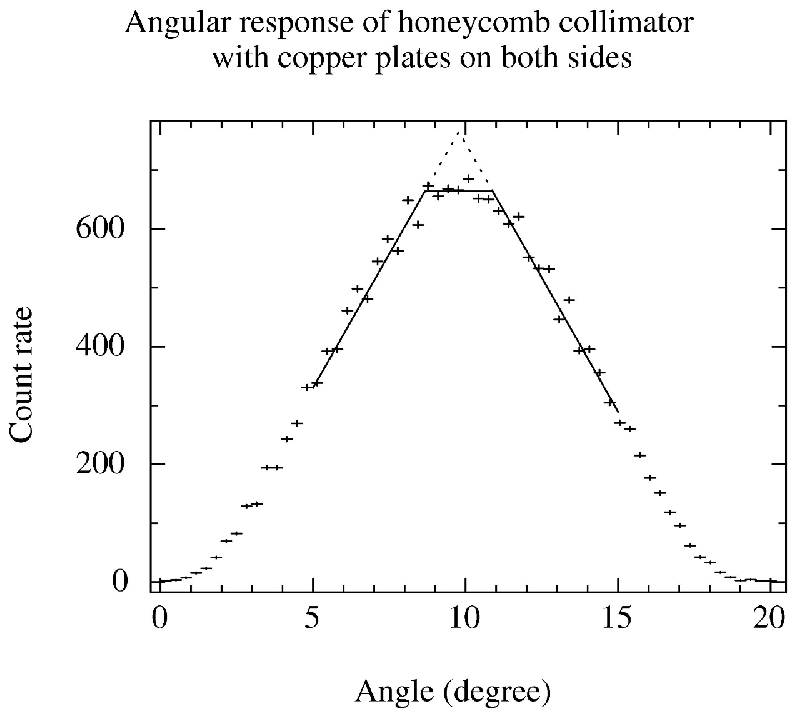}
\includegraphics[scale=0.35]{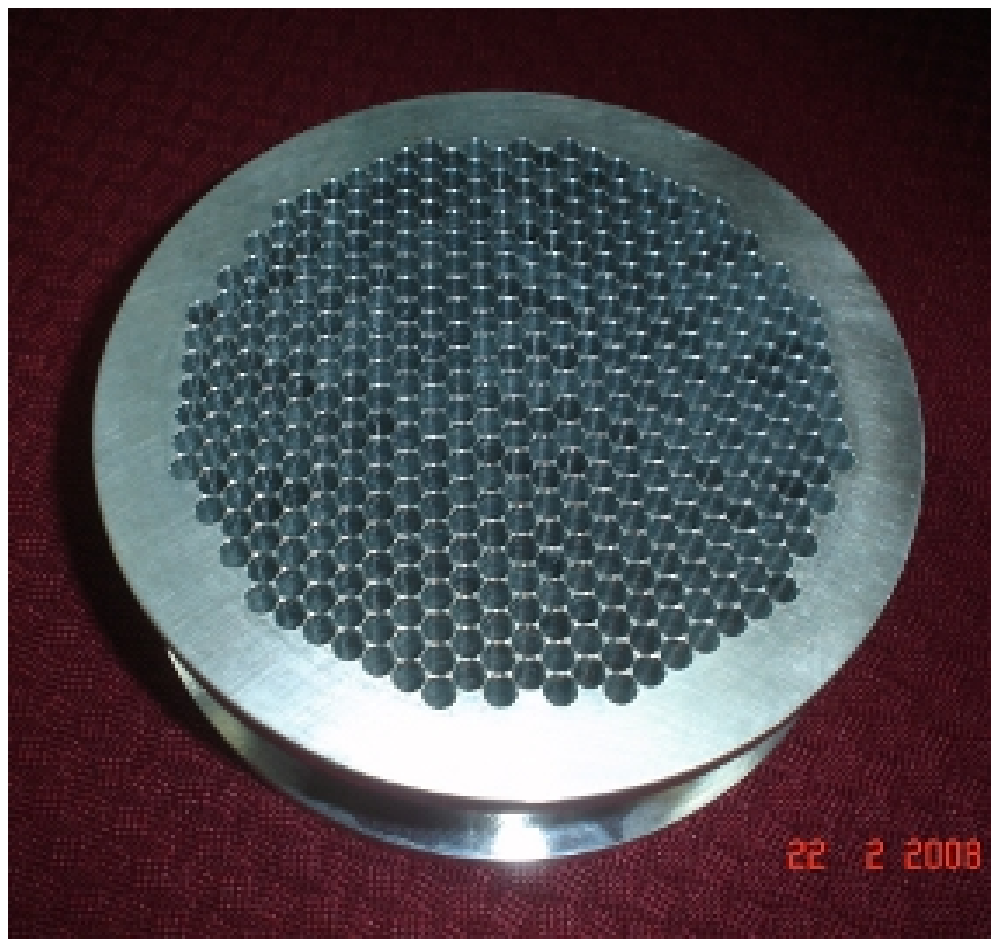}
\hfill \hfill \hfill
\caption{Collimators with honeycomb based and solid Al based construction.}
\label{f3}
\end{figure}

{\it Collimator:}
In order to compensate for inaccuracy in satellite pointing and to attain constant effective area, a collimator with a flat topped response is required.
For the Thomson Polarimeter, we have tested two collimator constructions which gives flat-topped response. The first construction uses aluminium honeycomb, sandwitched between copper plates with circular holes as shown in Figure \ref{f3}. The holes on the front plate has slightly smaller diameter than the holes on the back plate which leads to a flat topped response.  Since aluminium does not provide enough X-ray absorption above 15 keV, a silver coating of \mbox{$\sim$ 10 micron} thickness is provided on the aluminium surface to increase absorption. This construction has the advantage of small weight per unit area coverage. The second construction uses solid aluminium block with machined tapered holes to acheive a flat topped response. For the flight unit, a collimator with a flat topped response of 0.5$^\circ$ will be developed.

% Figure: Collimators- honey comb and solid Al

{\it Test Electronics:}
The different electronic units required for pulse processing and data acquisition from the detectors have been developed. This includes (1) HV distribution and Charge Sensitive Preamplifiers (2) Anti-coincidence counters and (3) Coincidence counters. We have also developed a general purpose  fixed-threshold pulse counter system capable of  acquiring four 8-bit channels simultaneously. An FPGA based Multi-channel Analyzer with ADC input and configurable pulse selection logic (coincidence, anti-coincidence etc.) is under development.

\section{Test and Calibration System}

% Figure: Rotary stage
\begin{figure}[h]
\centering
\includegraphics[scale=0.35]{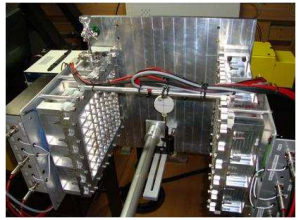}
\caption{The rotary stage with the detectors mounted.}
\label{f4}
\end{figure}

In order to test and calibrate the polarimeter, the following systems are at various stages of development. (1) A double crystal monochromator/polariser working in the energy range 2--60 keV to facilitate calibration of detectors over the entire energy band of interest. The total energy range will be covered using three pairs of crystals with d=1.15, 1.9 \& 3.1 A$^\circ$. The first crystal plate with d=3.1 A$^\circ$ has been fabricated, with the crystals parallelised to arc second level. The crystal plate has been tested in the energy range of \mbox{7--11 keV} and has FWHM of the beam $\sim$ 2 \% in this range. (2) A polarised source based on Thomson scattering of X-rays from a low Z scatterer. (3) A polarised source based on filtering out the highest energy (last 10 \%) photons from an electron beam X-ray generator (4) A rotary stage to simulate the spinning satellite platform during polarisation measurements (shown in Figure \ref{f4})

\section{Results}

% Figure: Results
\begin{figure}[h]
\centering
\includegraphics[scale=0.64]{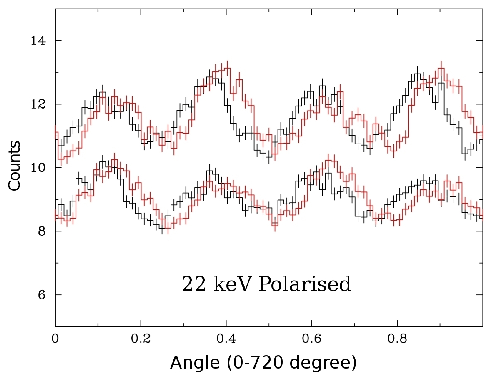}
\includegraphics[scale=0.6]{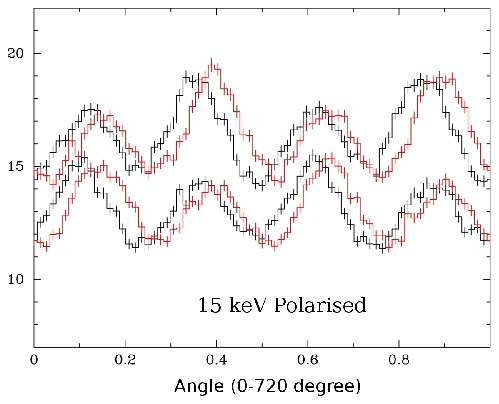}
\includegraphics[scale=0.6]{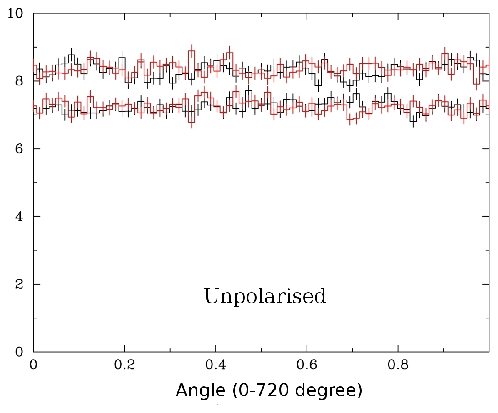}
\includegraphics[scale=0.6]{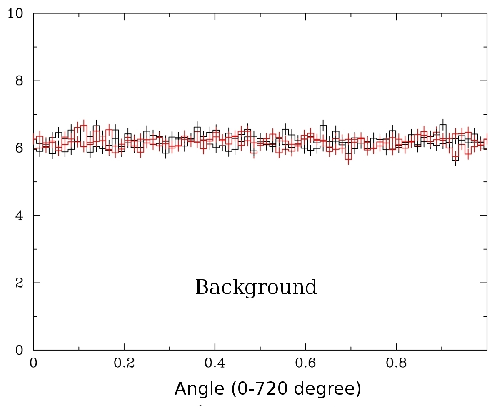}
\caption{Modulation obtained at different energies}
\label{f5}
\end{figure}

A laboratory model of the Thomson X-ray polarimeter with a Be scatterer and rectangular detectors has been completed and tested. Figure \ref{f5} shows the test results. In the 9--23~keV range, we obtained a modulation factor of up to 35 \%. The modulation factor is improved to about 45 \% when only one third of the detector is exposed, thereby reducing the opening angle. The experiment has been proposed for a small satellite mission of the Indian Space Research Organization (ISRO). 

\section*{Acknowledgements}
We thank members of the RAL and MES of Raman Research Institute for their excellent support in development of the instrument.

%%%%%%%%%%%%%%%%%%%%%%%%%%%%%%%%%%%%%%%%%%%%%%%%%%%%
\end{document}